# Adaptive evolution of hybrid bacteria by horizontal gene transfer

Jeffrey J. Power[1,♦], Fernanda Pinheiro[1,♦], Simone Pompei[1,♦], Viera Kovacova[1], Melih Yüksel[1], Isabel Rathmann[1], Mona Förster[1], Michael Lässig[1]*, Berenike Maier[1]*

[1] University of Cologne, Institute for Biological Physics, Köln, Germany

♦ Joint first authors.
* Corresponding authors. Email: mlaessig@uni-koeln.de, berenike.maier@uni-koeln.de.

**Abstract**

**Horizontal gene transfer is an important factor in bacterial evolution that can act across species boundaries. Yet, we know little about rate and genomic targets of cross-lineage gene transfer, and about its effects on the recipient organism's physiology and fitness. Here, we address these questions in a parallel evolution experiment with two *Bacillus subtilis* lineages of 7% sequence divergence. We observe rapid evolution of hybrid organisms: gene transfer swaps ~12% of the core genome in just 200 generations, and 60% of core genes are replaced in at least one population. By genomics, transcriptomics, fitness assays, and statistical modeling, we show that transfer generates adaptive evolution and functional alterations in hybrids. Specifically, our experiments reveal a strong, repeatable fitness increase of evolved populations in the stationary growth phase. By genomic analysis of the transfer statistics across replicate populations, we infer that selection on HGT has a broad genetic basis: 40% of the observed transfers are adaptive. At the level of functional gene networks, we find signatures of negative and positive selection, consistent with hybrid incompatibilities and adaptive evolution of network functions. Our results suggest that gene transfer navigates a complex cross-lineage fitness landscape, bridging epistatic barriers along multiple high-fitness paths.**

**Significance statement**

In a parallel evolution experiment, we probe lateral gene transfer between two *Bacillus subtilis* lineages close to the species boundary. We show that laboratory evolution by horizontal gene transfer can rapidly generate hybrid organisms with broad genomic and functional alterations. By combining genomics, transcriptomics, fitness assays and statistical modeling, we map the selective effects underlying gene transfer. We show that transfer takes place under genome-wide positive and negative selection, generating a net fitness increase in hybrids. The evolutionary dynamics efficiently navigates this fitness landscape, finding viable paths with increasing fraction of transferred genes.

**Introduction**

Horizontal gene transfer (HGT) plays an important role in bacterial evolution (1, 2), which includes speeding up the adaptation to new ecological niches (3, 4) and mitigating the genetic load of clonal reproduction (5, 6). On macro-evolutionary time scales, HGT occurs between bacteria of different species and also between bacteria and eukaryotes (1, 2, 7); these dynamics have very heterogeneous rates (8) and the permanent integration of transferred genes into regulatory networks is slow (9). An important mechanism of HGT is transformation, the active import and inheritable integration of DNA from the environment (10, 11). In this process, extracellular DNA binds to the surface of a recipient cell, is transported into the cytoplasm by the recipient's uptake machinery, and then integrated into its genome by homologous recombination.

The rate of transformation depends on multiple physiological and selective factors (12, 13). The efficiency of the DNA uptake machinery is a major determinant of the probability of transformation (11). At the level of recombination, this probability decreases exponentially as a function of the local sequence divergence, likely because nucleotide mismatches suppress sequence pairing at the initiation of the recombination step (14-16). As shown in a recent study, laboratory experiments can induce genome-wide transformation between close lineages (17). This study also describes the inhibition of gene uptake by restriction-modification systems, but fitness effects of gene transfer are not addressed. Other evolution experiments give evidence of complex selective effects of HGT (18-20). On the one hand, HGT can impose fitness costs that increase with genetic distance between donor and recipient (21). Observed costs include codon usage mismatch, reduction in RNA and protein stability, or mismatch of regulation or enzymatic activity in the recipient organism (22, 23), and can be mitigated by subsequent compensatory mutations (22, 24) or gene duplications (25). On the other hand, transferred genes can confer adaptive value, for example, resistance against antibiotics or shift of carbon sources (26, 27). These experiments elicit HGT of a few genes with a specific function, which are a confined genomic target of selection. However, we know little about positive and negative fitness effects of HGT on a genome-wide scale. In particular, it remains to be shown to what extent selection can foster genome-wide HGT across lineages, and where selective barriers significantly constrain recombination.

These questions are the subject of the present paper. We map genome-wide horizontal transfer between two model lineages (28) of *Bacillus subtilis,* a species with high genetic and phenotypic diversity (29). The two lineages are subspecies with an average sequence divergence of 6.8% in their core genomes, which is close to the species boundary: it is larger than typical diversities within populations and smaller than cross-species distances. We find pervasive HGT on a genome-wide scale, which covers hundreds of genes within about 200 generations and confers a repeatable net fitness increase to evolved hybrid organisms. Using Bayesian analysis, we jointly infer the effects of sequence diversity and selection on the observed transfer pattern. We find evidence for genome-wide positive and negative selection with significant hot- and cold spots in functional gene networks. We discuss the emerging picture of horizontal gene transfer navigating a complex cross-lineage fitness landscape.

**Results**

**Parallel evolution experiments.** Our experiments record HGT from the donor species *Bacillus subtilis subsp. spizizenii W23* to the recipient *Bacillus subtilis subsp. subtilis 168*. The two lineages (subspecies) share a 3.6 Mbp core genome with 3 746 genes. The average divergence of 6.8% in the core genome is small enough so that high-rate transfer is physiologically possible (30) and large enough so that transfer segments can reliably be detected in sequence alignments (SI Appendix, Fig. S1). In addition, there are unique accessory genomes of 0.4 Mbp in the donor and 0.6 Mbp in the recipient, which allow for non-orthologous sequence changes. Our evolution experiment consists of two-day cycles with a 6-step protocol shown in Fig. 1a. On day 1, recipient cells are diluted and cultured in liquid for 4.5 h, subject to UV radiation, and again diluted and inoculated onto an agar plate. On day 2, a single (clonal) colony is grown for 2.5 h in liquid culture, competence is induced and transformation in the presence of donor DNA takes place for 2 h, and then the culture is washed and grown overnight. The last step involves population dynamics with an exponential growth phase of about 4 hrs and a stationary growth phase of about 14 hrs. In this step, selection acts on HGT. In each run of the experiment, a recipient population evolves over 21 consecutive cycles, corresponding to about 200 generations. The full experiment consists of 7 replicate runs. We also perform 3 control runs without donor DNA, which are mutation-accumulation experiments under the same population-dynamic conditions as the primary runs. Our laboratory protocol is designed to study HGT at controlled levels of donor DNA and of induced competence. This differs from natural environments, where both factors are subject to environmental fluctuations, regulatory tuning, and ecological interactions between lineages (11, 31).

**Gene transfer is the dominant mode of evolution.** The experiments reveal fast and repeatable genome evolution that is driven predominantly by HGT. To infer these dynamics, we perform whole-genome sequencing after cycles 9, 15, and 21, and align the evolved genomes to the donor and the ancestral recipient genome (SI Appendix, Fig. S1). The genome-wide pattern of sequence evolution in hybrids is displayed in Fig. 1b and Fig. S2. Transferred genome segments (green bars) are seen to be distributed over the entire core genome of the recipient. We find HGT by orthologous recombination of genome segments at an approximately constant rate of about 11 genes/hr (counted over the transfer period of 2 hrs/cycle) in all replicate runs. These changes accumulate to an average of about 100 transferred segments per replicate, which cover 12% ($\pm$ 3%) of the core genome and affect 500 ($\pm$ 133) core genes (Fig. 1c). Thus, HGT by orthologous recombination causes sequence evolution at an average rate of 100 bp/generation (counted over all cycles). This drastically exceeds the mutation rate of $2 \times 10^{-3}$ bp/generation determined by a fluctuation assay in the absence of radiation (SI Appendix), as well as the corresponding rate recorded under UV treatment, $3 \times 10^{-1}$ bp/generation (corresponding to an average of 55 de novo mutations per replicate, Table S1). The HGT rate is also higher than found previously in *Helicobacteri pylori* (17), despite the substantially larger average sequence divergence between donor and recipient.

Besides orthologous recombination, we find insertions from the donor accessory genome (average 5 genes per replicate), deletions (average 31 genes per replicate) including the deletion of the mobile element ICEBs1 (in all replicates but not in the control runs, indicating clearance by HGT), and multi-gene duplications (192 genes in one replicate); this data is reported in Table S2. Thus, the by far dominant mode of genome evolution is HGT by orthologous recombination (simply called HGT in the following).

**Transfer depends on segment length and sequence similarity.** Transferred segments have an average length of 4200 bp, contain an average of 5.1 genes, and have an approximately exponential length distribution (Fig. 1d), in agreement with previous results in *Streptococcus pneumoniae* (32) and *Haemophilus influenza* (33). These statistics show ubiquitous multi-gene transfers with no sharp cutoff on segment length.

Fig. 1e displays the distribution of donor-recipient sequence divergence, $d$, in transferred segments (recorded in 100 bp windows around the inferred recombination start site). By comparing this distribution to the divergence distribution recorded in randomly positioned 100 bp windows, we infer a transfer rate $u(d)$ that decreases exponentially with increasing $d$; see SI Appendix and Fig. S3 for details. This result is in agreement with previous work (14, 15). In our system, the dependence on sequence similarity already generates a significant variation of transfer rates independently of selection on specific genes. About 16% of potential recombination start sites in the core genome have a local divergence < 5%, leading to a transfer rate higher than 3-fold above the genome average; other 3% have a divergence > 15% and a transfer rate lower than 9-fold below the average (Fig. S3). We conclude that the transfer rate is significantly modulated by the local efficiency of the recombination machinery, but transfer is physiologically possible at the vast majority of genomic sites.

**Gene transfer is broadly distributed.** Next, we ask how HGT is distributed over replicate runs and genome loci. For a given gene, the transfer frequency, θ, is defined as the fraction of runs in which that gene is hit by HGT. Fig. 1f shows the histogram of transfer frequencies evaluated over the entire core genome. We find 60% of core genes are transferred in at least one of 7 runs, and 1.7% of core genes are replaced in 4 or more runs. This is consistent with the broad genomic distribution of HGT shown in Fig. 1b. We find only few exceptions from this pattern. First, two hot spots of about 8 kbp and 16 kbp show strong enhancement of HGT, with more than 50% of their genes repeatably swapped in all replicates (Fig. 1f, SI Appendix). Both hot spots encode functional differences between donor and recipient, which are discussed below. To assess their statistical significance, we compare the observed multiple-hit statistics with a null model that is obtained by simulations of a positionally scrambled HGT dynamics with local rates $u(d)$ (SI Appendix, Extended Data Fig. S3). Both hot spots turn out to be highly significant ($P < 2 \times 10^{-3}$); that is, their multiple-hit statistics cannot be explained by local sequence divergence alone. Second, there are two genomic cold spots, which are extended segments where HGT is repeatably suppressed in all replicates. Both cold spots are about 50 kb long and are significant deviations from the null model; that is, the absence of transfer is unlikely to be caused by local sequence divergence alone ($P < 0.04$, Fig. S4, SI Appendix).

Analyzing HGT by gene ontology (GO) leads to a similarly broad distribution (SI Appendix, Table S3). In most GO categories, the observed average transfer frequencies are consistent with the null model; only essential genes show significantly enhanced transfer ($> 30\%$, $P < 0.03$). Taken together, most genomic loci and most functional classes are accessible to evolution by cross-lineage HGT, allowing the evolution of complex hybridization patterns.

**Gene transfer generates a net fitness gain.** We measure the selection coefficients of evolved hybrids and of control populations in competition with ancestral recipient cells. Selection is evaluated separately for the exponential and stationary growth phases, which are two consecutive selection windows in our population dynamics following HGT (Fig. 2a). The stationary phase covers most of the population dynamics interval (14/18 hrs). We find a repeatable fitness increase of about 5% in this phase, which is statistically significant ($P <$ 0.03, Mann-Whitney U-test) and repeatable across replicate populations. This net fitness increase signals adaptive evolution by HGT; that is, the uptake of donor genes does more than just repair deleterious mutations caused by UV radiation. Consistent with this interpretation, the control populations evolving by mutation accumulation show no comparable fitness increase (Fig. 2a).

In the exponential growth phase, the replicate-average fitness remains constant, so that the total balance in average fitness is dominated by the gain in the stationary phase. The control populations decline in average fitness in both phases, as expected for mutation accumulation under UV treatment. Evolved hybrids acquire a large exponential-phase fitness variation across replicate runs (Fig. 2a). This pattern indicates that complex fitness landscapes with positive and negative components govern the evolution of hybrid organisms; the shape of such landscapes will be further explored below.

**Gene transfer has genome-wide selective effects.** How are the strong fitness effects of HGT compatible with the ubiquity of transfer across the core genome? In particular, is adaptive evolution by HGT limited to few genomic loci with large effects or can we identify a genetic basis of multiple genes with potentially smaller individual effects? To establish a link between genotype and fitness, we first analyze the multiple-hit statistics of transfers in comparison to the null model of scrambled transfers with rates $u(d)$ (SI Appendix, Fig. S3). By simulations of the null model, we obtain a transfer probability $p_0$ for each gene that depends only on the donor-recipient mutation pattern in its vicinity. This probability accounts for the fact that local sequence divergence affects homologous recombination. The null model is neutral with respect to gene function; i.e., all genes in loci with the same $p_0$ have the same multiple-hit probability distribution $P_0(\theta|p_0) = B(7, 7\theta, p_0)$, which is a binomial distribution with expectation value $p_0$ (SI Appendix). To quantify target and strength of selection, we use a minimal model with selectively enhanced (i.e., adaptive) transfer probabilities $p_+ = \varphi_+ p_0$ in a fraction $c$ of the core genes and reduced probabilities $p_- = \varphi_- p_0$ in the remainder of the core genome. This mixture model generates a multiple-hit distribution

$$Q(\theta| p_0) = c\, B(7, 7\theta, \varphi_+ p_0) + (1 - c)\, B(7, 7\theta, \varphi_- p_0). \qquad [1]$$

Importantly, the model jointly captures neutral and selective variation of HGT rates, properly discounting physiological effects from the inference of selection. Comparing the observed distribution of multiple hits, $\hat{Q}(\theta|p_0)$, and the null distribution shows an excess count of no-hit ($\theta = 0$) genes and of multiple hits ($\theta \geq 3/7$). Importantly, this pattern is observed in all $p_0$ classes; i.e., independently of the local sequence similarity. The deviations from the null model are strongly significant ($P < 10^{-20}$, SI Appendix), and we ascribe them to selection by HGT. The selection mixture model, with parameters $c = 0.2\, [0.1, 0.4]$, $\varphi_+ = 1.9\, [1.6, 2.4]$, and $\varphi_- = 0.75\, [0.6, 0.84]$ (maximum-likelihood value, confidence interval in brackets) estimated from a Bayesian posterior distribution, explains the observed multiple hits (Fig. 2c) and infers $40\, [20, 60]$ % of the observed transfers to be adaptive (SI Appendix, Fig. S6). These results suggest that HGT is a broad target of positive and negative selection in hybrid organisms, in tune with the experimental results (Fig. 2a).

**Frequently transferred genes tend to be upregulated and positively selected.** Next, we characterize the effect of HGT on gene expression. Using whole-genome transcriptomics data, we compare gene expression in evolved strains and their ancestor (Fig. 3, Fig. S4). The overall distribution of log$_2$ fold changes of RNA levels $\Delta R$ in the entire genome is balanced and similar to that observed in the control runs, as expected for viable cells (raw data are reported in Table S4). To map correlations between HGT and expression, we partition genes into a class with low transfer frequency, $0 \leq \theta \leq 3/7$, and a class with high transfer frequency, $4/7 \leq \theta \leq 1$. Within each class, we further partition genes by their ancestral lineage (recipient: R, donor: D). We observe that genes with low transfer frequency are not substantially affected in their average expression level. In contrast, genes with high transfer frequency show upregulation in hybrids, which is strongest for genes hit by HGT. A priori, upregulation can compensate for reduced translational or functional efficiency, or it can signal an enhanced functional role of the affected genes. In the first case, we would expect a signal for transferred (D) genes independently of their $\theta$ class. The fact that we find upregulation specifically for genes with high transfer frequency but not for others may point to an enhanced functional role. This is consistent with our inference of selection: in the high-transfer class, 88% of the genes are inferred to be under positive selection (Methods).

**Selection on gene transfer in functional networks.** To link selection to cellular functions, we map the transfer patterns onto gene networks of the recipient organism. Specifically, we use complex, multi-gene operons as units of genomic and functional organization above the level of individual genes. To explain the observed transfer statistics, we use a minimal cross-lineage fitness landscape of the form

$$F(q) = aq + bq(1-q), \qquad [2]$$

which quantifies the selective effect of an operon as a function of the transfer fraction $q$, relative to the ancestral ($q = 0$) state (Fig. 4a). The linear term describes directional selection on HGT, the quadratic term intra-network epistasis. Specifically for $b < 0$, this term captures hybrid incompatibilities within operons, leading to a fitness trough at intermediate $q$ and a rebound at larger values of $q$.

Fig 4b shows the distribution of the average transfer frequency $\bar{\theta}$ for operons containing at least 7 genes (bars denote averages over all genes in a given operon); this threshold captures the 56 most complex operons. For a given operon, $\bar{\theta}$ equals the fraction of its HGT-affected genes, $q$, averaged over all replicate runs. The corresponding distribution of $\bar{\theta}$ under neutral evolution, which is again obtained by simulations of positionally scrambled transfers (SI Appendix), discounts all HGT correlations between genes due to their spatial clustering in a common operon. To compare data and neutral model, we omit the two hotspots of HGT identified above, which contain operons discussed below. In the remaining data, we find a statistically significant excess of operons with small observed transfer frequency (46% of counts in the data with $\bar{\theta} < 0.1$ vs. 31% in the null model), signaling selection against HGT at the network level ($P = 0.01$., SI Appendix). As shown by simulations of the HGT dynamics in the fitness landscapes of Fig. 4a (SI Appendix), the observed $\bar{\theta}$ distribution in complex operons is consistent with hybrid incompatibilities ($b < 0$) or with negative directional selection ($a < 0$) (Fig. 4c). In Fig. 4d, we show two examples of operons with suppressed transfer. One operon is involved in pectin utilization, while the other one is involved in inositol utilization (Table S5). Both operons have none of their genes replaced in any of the 7 replicates, which is consistent with negative directional selection ($a < 0$). Below each plot, we also display the Protein-Protein Interaction (PPI) links within the operon, which may be indicative of functional links between the operon genes (34).

The two genomic hotspots of HGT harbor operons with high average transfer frequency (Fig. 4b,e). The *leu* operon has $\bar{\theta} = 0.84$ and 100% of its genes are coherently transferred in 3 replicate runs. The *leu* operon confers the ability to grow without external leucine supply. Evolved hybrids gain this function, which is present in the donor but absent in the ancestral recipient (SI Appendix). Uptake of *leu* has been observed previously in transformation essays under leucine starvation (36); our stationary-phase protocol may generate selection for leucine synthesis. The *eps* operon, which contains 13 genes, has $\bar{\theta} = 0.64$ and up to 100% transferred genes. This operon is important for biofilm formation, a function that is strongly impaired in the recipient (35) but potent in the donor. Both operons show an enhancement of $\bar{\theta}$ that deviates significantly from the neutral null model and is consistent with positive directional selection ($a > 0$) (Fig. 4e, SI Appendix). This signals adaptive evolution by HGT at the level of gene networks, generating coherent transfer of its constituent genes. Additional examples of operons with strongly enhanced $\bar{\theta}$ are shown in Fig. S7.

**Discussion**

In this work, we integrate experimental evolution and evolutionary modeling to map effects of horizontal gene transfer on genome dynamics, gene expression and fitness. We show that genome-wide cross-lineage HGT is a fast evolutionary mode generating hybrid bacterial organisms. This mode is repeatably observed in all replicate runs of our experiment. After about 200 generations, *Bacillus* hybrids have acquired about 12% donor genes across the entire core genome, but coherently transferred functional gene networks reach up to 100% transferred genes in individual runs.

Despite its broad genomic pattern, the HGT dynamics is far from evolutionary neutrality. Evolved hybrids show a substantial fitness increase compared to the ancestral strain in stationary growth, which occurs repeatably across all replicates. Hence, HGT does more than just repairing deleterious mutations, which are caused by UV radiation in our protocol: it carries a net adaptive benefit. The adaptive dynamics has a broad genomic target; we infer some 40% of the observed transfers to be under positive selection. In addition, we find positive and negative selection on HGT in a range of functional gene networks. We conclude that in our system, unlike for uptake of resistance genes under antibiotic stress (26, 27), HGT does not have just a single dominant target of selection. Rather, evolution by orthologous recombination appears to tinker with multiple new combinations of donor and recipient genes, using the combinatorial complexity of hundreds of transferred genes in each replicate run. This picture is consistent with the substantial fitness variation across replicates, which is most pronounced in the exponential growth phase.

To capture HGT under genome-wide selection, we introduce the concept of a cross-lineage fitness landscape. Such landscapes describe the fitness effects of orthologous recombination between two (sub-)species, starting from the unmixed genomes as focal points of *a priori* equal rank. This feature distinguishes cross-lineage landscapes from empirically known landscapes for mutation accumulation within lineages, e.g. for antibiotic resistance evolution, most of which have a single global fitness peak. Here we begin to map building blocks of cross-lineage landscapes (Fig. 4). These include positive selection for uptake of gene networks, well as selection against transfer in genes and operons, suggesting that hybrid incompatibilities between donor and recipient and epistasis may play a role in the observed HGT dynamics. Future massively parallel experiments will allow a more systematic mapping of cross-species fitness landscapes, which combines these building blocks into a systems picture. Such experiments will also show how different compositions of donor DNA and different selection pressures applied in the experimental cycle modulate the HGT dynamics on cross-lineage landscapes.

Our observation of fast HGT under broad selection suggests that evolution navigates the cross-lineage landscape in an efficient way. That is, deep cross-lineage fitness valleys must be sparse enough for evolution by HGT to find viable paths of hybrid evolution with increasing transfer fraction. Stronger and more ubiquitous fitness barriers at larger donor-recipient distances or larger transfer fractions may eventually halt gene uptake, but we have yet to see where the limits of genome-wide HGT are.

Two key features of our experiments enable these dynamics. First, choosing a donor-recipient pair with a sequence divergence close to the species boundary generates more, often subtle functional differences between orthologous genes – that is, more potential targets of selection on HGT – than closer pairs or single populations. Second, permissive population dynamics with recurrent bottlenecks allow the (transient) fixation of deleterious recombinant genomes, which can bridge fitness valleys of hybrid incompatibilities and act as stepping stones for subsequent adaptation. Together, our results suggest that laboratory evolution by HGT can become a factory for evolutionary innovation. An exciting perspective is to use cross-lineage HGT together with artificial evolution in order to engineer functional novelties.

**Figures**

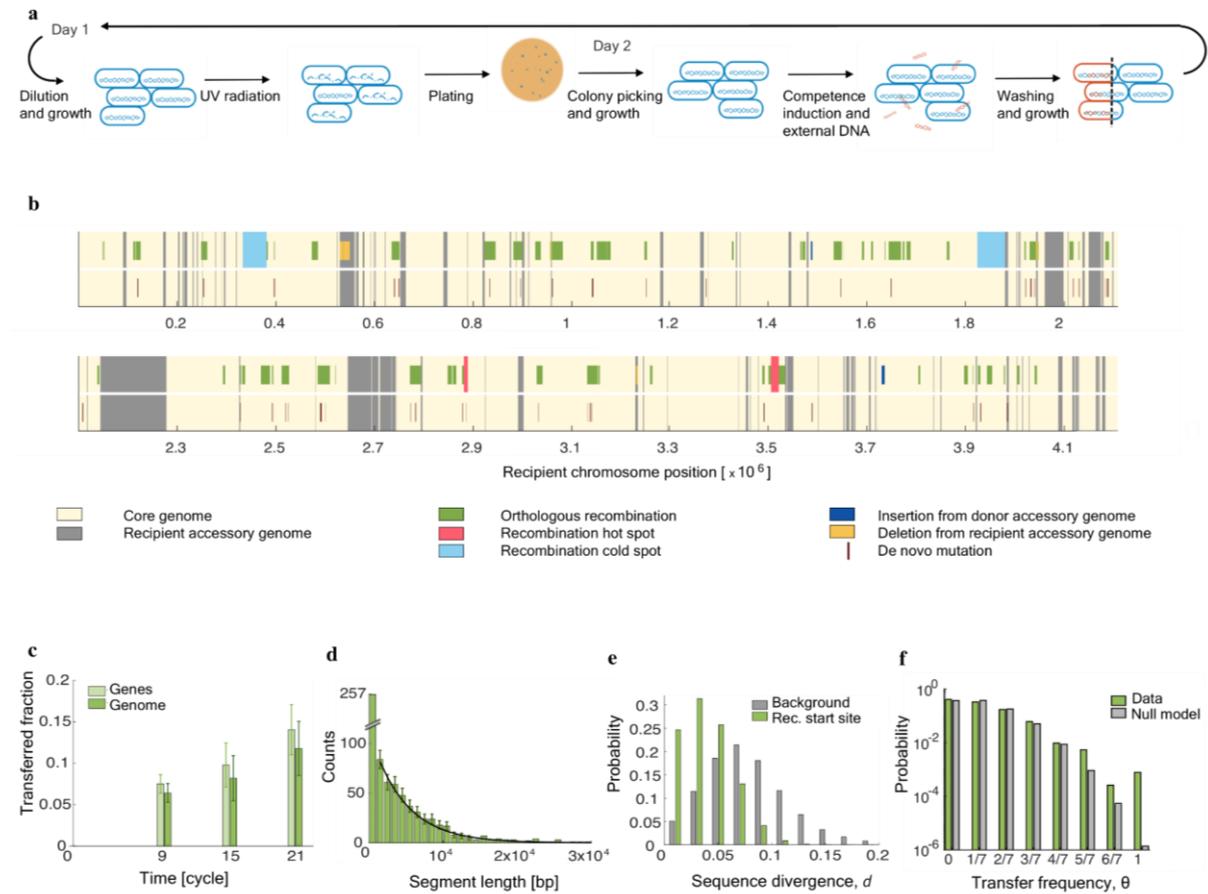

**Fig. 1. Evolution of *B. subtilis* hybrids.** **(a)** Experimental design. A single two-day (10 generation) cycle contains 6 steps, including two growth phases with irradiation and horizontal gene transfer (HGT), respectively. **(b)** Transferred segments (green) are broadly distributed over the recipient core genome (light shading) with two hot spots (red) and two cold spots (blue). De novo mutations occur throughout the recipient genome, and there are a few deletions from the recipient genome (orange). Data are shown for run 4; see Fig. S3 and Tables S1 and S2 for data from all runs. **(c)** Time-dependence of HGT. Fraction of core genes (light green) and of core genome (dark green) affected by HGT after 9, 15, and 21 cycles (mean and standard deviation across the 7 parallel runs). **(d)** Length distribution of transferred genome segments (bars: count histogram, black line: exponential fit for segments > 1000 bp). **(e)** Inferred distribution of donor-recipient sequence divergence, $d$, in 100 bp windows around the recombination start site of transferred segments (green) and in scrambled 100 bp windows (gray). **(f)** Distribution of the transfer frequency per gene, $\hat{Q}(\theta)$, in the 7 parallel runs, counted across the recipient core genome (green); corresponding distribution $P_0(\theta)$ from simulations of the null model (gray).

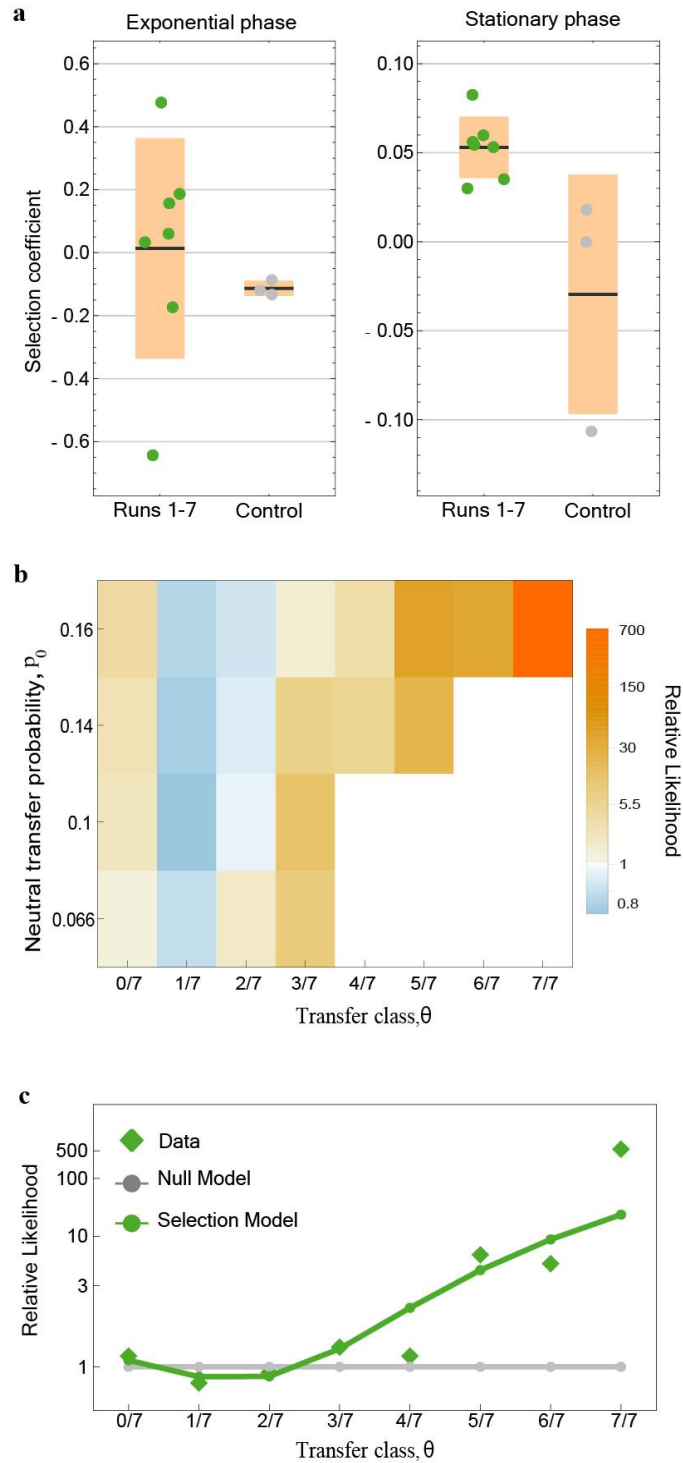

**Fig. 2. Selective effects of horizontal gene transfer.** (a) Transfer affects hybrid fitness. Mean selection coefficient of hybrids compared to the ancestral recipient population in the exponential phase and in the stationary phase (colored dots: data from individual runs, bars and boxes: mean and standard deviation over all 7 runs); control data of runs without donor DNA (gray). (b) Genome-wide selection on transfer. The relative likelihood of transfer, $\hat{Q}(\theta|p_0)/P_0(\theta|p_0)$, is shown for genes binned by transfer class ($\theta = 0/7, 1/7, ..., 7/7$) and by the neutral transfer probability $p_0$, which depends on sequence similarity. (c) The relative likelihood of transfer aggregated over $p_0$ bins, $\hat{Q}(\theta)/P_0(\theta)$, is shown together with the corresponding likelihood ratio $Q^*(\theta)/P_0(\theta)$ of the maximum-likelihood selection model (green circles); see SI Appendix. Relative to the neutral null model (gray baseline), the selection model has an enhanced transfer probability ($p_+/p_0 = 1.9$) in a fraction $q = 0.2$ of the recipient core genes and a reduced probability ($p_-/p_0 = 0.75$) in the remainder of core genes.

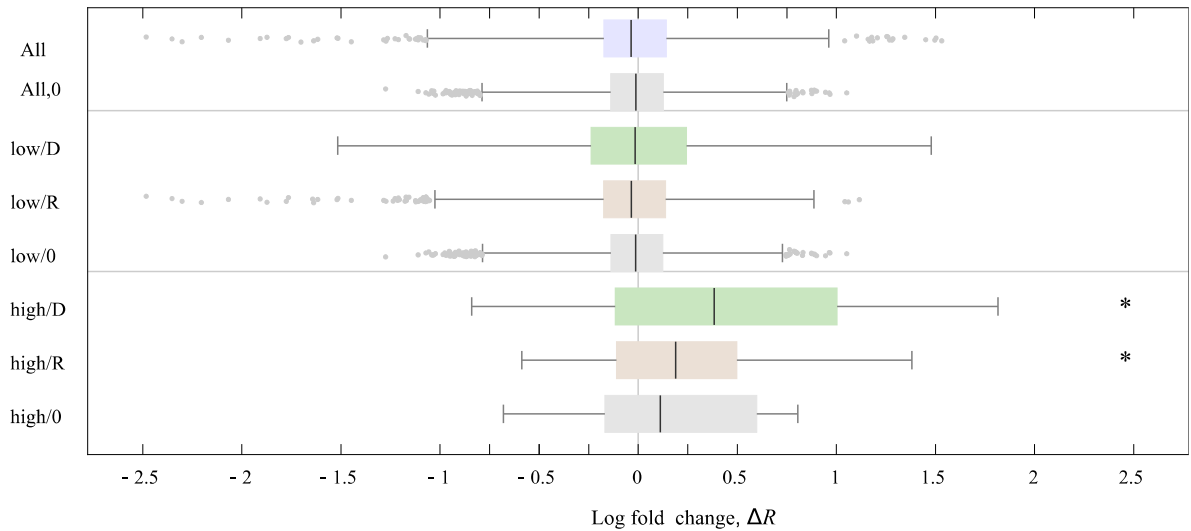

**Fig. 3. Gene expression in hybrids.** Log$_2$ RNA fold changes, $\Delta R$, with respect to the ancestral recipient (whisker plots, blue line: mean, box: 1$^{st}$ and 3$^{rd}$ quartiles, bars: 99% percentiles, dots: outliers) are shown for different gene classes: all genome, genes with low transfer frequency ($0 \leq \theta \leq 3/7$), and genes with high transfer frequency ($4/7 \leq \theta \leq 1$); cf. Fig. 1f. In each gene class, we show $\Delta R$ whisker plots separately for non-transferred recipient genes (R) (red) and for transferred orthologous donor genes (D) (green), together with the corresponding changes in the control experiments without donor DNA (0) (grey). For example, a gene subject to HGT in replicates 1, 2, 5 is in the low-transfer class; its $\Delta R$ values from replicates 1, 2, 5 (3, 4, 6, 7) contribute to the low/D (low/R) statistics. Asterisks mark statistically significant changes of the average $\Delta R$ for upregulation of high/D and high/R genes ($P < 10^{-3}$, T-test) compared to the ancestor.

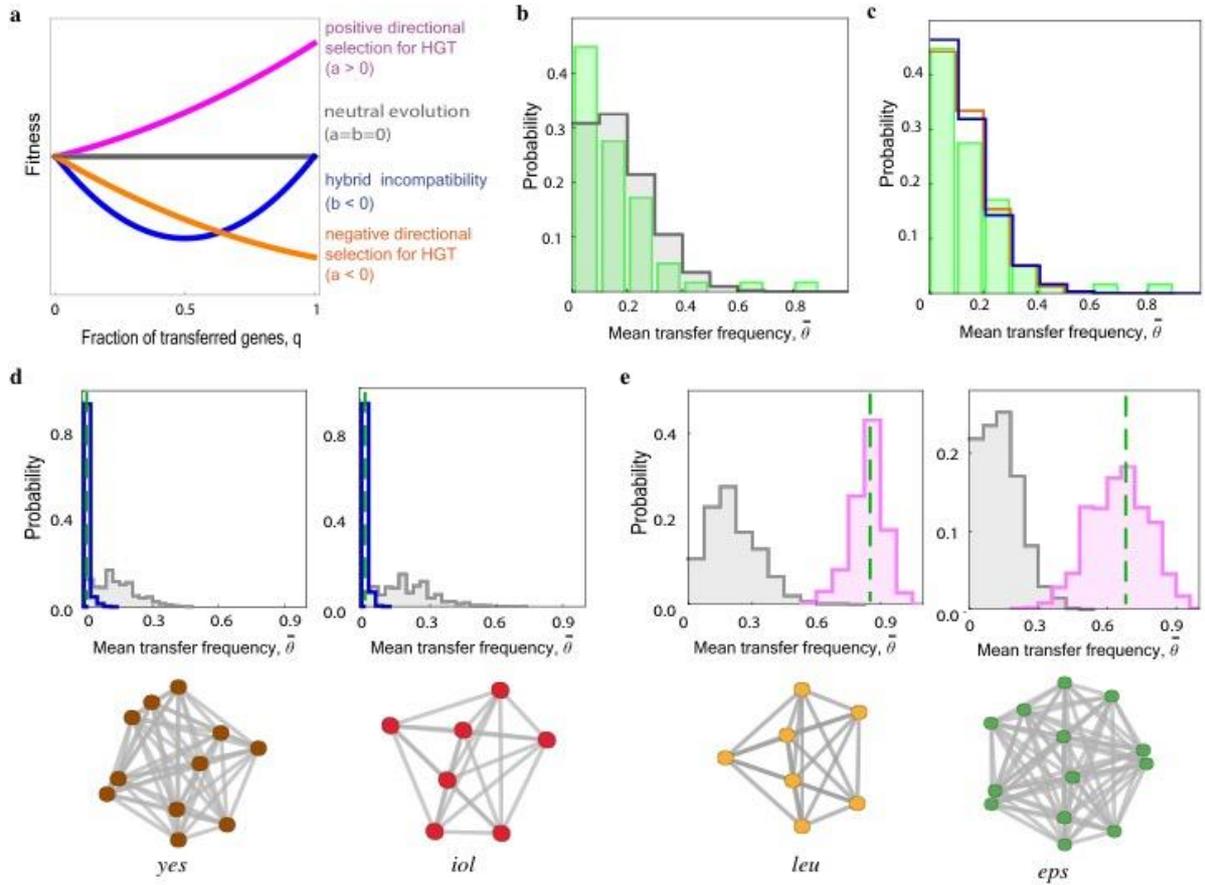

**Fig. 4. Cross-lineage fitness effects in gene networks.** (**a**) Cross-lineage fitness landscapes given by eq. [1] with hybrid incompatibilities ($b < 0$, blue), predominantly negative directional selection ($a < 0$, orange), and predominantly positive directional selection ($a > 0$, magenta). (**b-c**) Distribution of the observed average transfer frequency, $\bar{\theta}$, for recipient operons containing at least 7 genes (green). Corresponding distributions (b) for the neutral null model (gray) and (c) for fitness models given by eq. [1] with hybrid incompatibilities ($b < 0$, blue) and with negative directional selection ($a < 0$, orange). (**d**) Examples of low-transfer operons, *yes* and *iol*. Observed values of $\bar{\theta}$ (dashed lines) are reduced in comparison to the neutral null model (gray) and consistent with negative directional selection ($a < 0$, blue); see SI Appendix. (**e**) Examples of high-transfer operons, *leu* and *eps*. Observed values of $\bar{\theta}$ (dashed lines) are enhanced in comparison to the neutral null model (gray) and consistent with positive directional selection ($a > 0$, magenta). The deviation from the neutral model is statistically significant (*leu*: $P < 10^{-3}$, *eps*: $P < 10^{-3}$; SI Appendix). PPI links for these operons are shown below (34). Simulations of transfer under selection use the fitness landscape of eq. [2] with selection coefficients $a, b$ listed in SI Appendix.

**SI Appendix**

The SI Appendix describes experimental procedures, sequence analysis, and evolutionary analysis of HGT.

**Acknowledgements**. We acknowledge discussions with M. Cosentino Lagomarsino and A. de Visser. This work has been partially funded by Deutsche Forschungsgemeinschaft grant CRC 1310.

**Author Contributions.** First authors: JJP developed, performed, and analyzed the core experiments; FP and SP developed and performed evolutionary analysis. JJP, FP, SP, MY, ML, and BM designed research; JJP, MY, IR, and MF performed the experiments; VK and JJP processed sequence data; VK and MY processed transcriptomics data; ML and BM wrote the article; all authors discussed and approved the article.

# SI Appendix

## 1. Experimental procedures

**Strains, Media, and Growth.** The donor strain used was 2A9 (*B. subtilis subsp. spizizenii W23*). All other strains used in this study were derived from strain BD630 (*B. subtilis subsp. subtilis 168 his leu met*). *B. subtilis* strains were grown either in liquid LB medium or in competence medium (1), supplemented with 0.5 % glucose, 50 µg/ml L-histidine, L-leucine, L-methionine, and kanamycin (5µg/ml), or spectinomycin (100 µg/ml) at 37°C. Growth was monitored by measuring the $OD_{600}$ on an Infinite M200 plate reader (Tecan, Männedorf, Switzerland). Competent cells were prepared as described previously and transformed by following the standard protocol (1).

In *B. subtilis*, natural competence for transformation develops at high cell densities. Competence development is regulated by an extended network of regulatory proteins (2). This regulatory network is a large mutational target and mutations are likely to affect the probability of competence development and thus HGT. Since competence development is not subject of this study, we circumvented this problem by setting the expression of the master regulator for competence, *comK*, under the control of an IPTG-inducible promoter. The recipient strain Bs166 (*his leu met amyE::P$_{hs}$comK(spc) comK::kan*) was generated by transforming strain BD3836 (3) with genomic DNA from strain Bs075 (4). For competition experiments, we generated the reporter strain Bs175 (*his leu met amyE::P$_{hs}$comK(spc) comK::kan lacA::PrrnE-gfp (erm)*) as follows. *PrrnE-gfpmut2* was amplified from genomic DNA of Bs139 (4) using primers MeY87 (5'- AAGGAATTCCGGATCCGAAGCAGGTTAT-3') and MeY88 (5'- GGACTAGTTTATTTGTATAGTTCATCCATGC -3'). Both the PCR-fragment and the donor plasmid pBS2E (obtained from the *Bacillus subtilis* Stock Center) were double digested with EcoRI and SpeI. The PCR-fragment was ligated into the linearized plasmid, after dephosphorylation of the 5'-sticky ends of pBS2E. The ligation product pBS2E_*PrrnE-gfpmut2* was then transformed into Bs166 to generate Bs175.

**Whole Genome Sequencing.** Clonal genomes of ancestral and evolved populations were obtained using next generation sequencing (NGS) methods, in particular Illumina HiSeq. Samples were prepared by growing a frozen culture overnight on an LB Agar plate at 37° C, 5% $CO_2$. The subsequent day, an individual colony was selected and grown overnight in competence medium. A 2 mL aliquot of that culture was pelleted at 16.7 x g for 3 min, decanted, and then frozen at -20° C. Additionally, a 1 mL aliquot of the overnight CM culture was mixed with DMSO (10% v/v) and stored for reference at -80° C. Genomic DNA was isolated from the frozen pellet using the Qiagen Dneasey Blood & Tissue Kit (Hilden, Germany) according to the manufacturer's instructions. A small aliquot of the Isolated DNA was run on a 1% agarose gel with a 1 kb plus DNA Ladder (Thermo Scientific) to check for degradation. Non-degraded samples were sent to GATC Biotech (Konstanz, Germany) for NGS. Sequencing was performed on an Illumina HiSeq 3000/4000 system with 150 bp paired reads and an average depth of >500.

**RNA sequencing.** Clonal transcripts of ancestral and evolved populations were obtained using next generation sequencing (NGS) methods, in particular Illumina HiSeq. Samples were prepared by growing a frozen culture overnight on an LB Agar plate at 37° C, 5% $CO_2$. The subsequent day, individual colonies were selected and grown in 1 ml competence medium. After 2.5 h growth, the cultures were diluted to an 0.1 OD in competence medium and were incubated for 3 h. Total RNA was isolated by using the Qiagen RNeasy Protect Bacteria Mini Kit (Hilden, Germany) according to the manufacturer's instructions. Samples were sent to Cologne Genomic Center (Cologne, Germany) for NGS. Prior to sequencing ribosomal RNA was depleted from the samples by using the Illumina Ribo-Zero rRNA Removal Kit (San Diego, USA). Sequencing was performed on an Illumina HiSeq system with 75 bp paired reads and on average 5 million reads per sample.

**Determination of mutation rate.** Spontaneous mutation rates to nalidixic acid resistance were determined using a fluctuation assay (5). Twenty parallel cultures grown overnight in LB medium were inoculated in competence medium with a starting $OD_{600}$ 0.01 and grown for 6 h. Subsequently, each culture was plated on LB plates containing 15 µg/ml nalidixic acid and incubated at 37°C. After 24 h, the number of colonies per plate was determined. The total number of cells was determined by plating dilutions of three parallel cultures on LB agar plates. The mutation rate was calculated with the MSS-Maximum-Likelihood Method (6). These experiments were repeated at least three times. The mutation rate was determined to be $(5.04 \pm 3.82) \times 10^{-10}$ per nucleotide per generation.

**Evolution Experiment.** The evolution experiment was composed of continuous repetitions of a two-day cycle, consisting of six steps: dilution, radiation, plating, colony selection and regrowth, competence induction and addition of extracellular DNA, washing and overnight growth (Fig. 1a). Seven replicate ancestral clones of the recipient strain $\Delta comK$ $comK_{ind}$ (Bs166) were used for transformation with genomic DNA from *B. subtilis* W23 and three clones were used for the control experiments. Initially, all strains were grown overnight in 1 mL LB medium at 37°C, 250 rpm.

Overnight cultures were diluted $1:10^3$ in fresh competence medium and grown for 4.5 h (37°C, 250 rpm) in a 24-well microtiter plate (1 mL, final OD ≈ 0.25; Greiner Bio-one). Microtiter plates were covered with rayon film adhesive covers (VWR). Cultures were radiated in the microtiter plate (without cover) in a 600 $Jcm^{-2}$ UV-C light chamber (Bio-Link BLX-E crosslinker). We included the irradiation step to damage the chromosomal DNA hypothesizing that donor DNA was used as a template for DNA repair as proposed earlier (7). A 100 µL aliquot of each radiated culture was diluted $1:10^4$, plated onto a 10 cm LB-agar plate, and incubated overnight at 37°C, 5% $CO_2$. A random colony was picked from each plate using a 200µL pipette tip, mixed into 1 mL fresh competence medium, and grown for 2.5 h (OD ≈ 0.25). IPTG [600µM] was added to each culture along with genomic DNA equivalent to two genomes per cell. Genome equivalents were calculated assuming 1 bp = 650 Da and a culture density of $10^8$ CFU/mL. After growing induced cultures for an additional 2 h (37°C), each culture was washed twice (16,800xg, 1 mL competence medium) and grown overnight (37°C, 250 rpm). This completed one cycle and was repeated, starting again with dilution. Samples were frozen for later analysis every second cycle, starting with cycle 3. For each culture, a 500µL mixture with DMSO (10% v/v) was stored at -80°C.

**Gain-of-function experiments.** Our statistical analysis identifies two HGT hot spots (section 3, Fig. S3). These hot spots contain operons encoding cellular functions that are present in the donor but absent in the recipient strain: the *leu* operon enables leucine synthesis, and the *eps* operon is important for biofilm formation. (i) We tested the gain of leucine synthesis in evolved hybrid strains (cycle 21) that have acquired the *leu* operon by growing the hybrids overnight in defined medium without leucine. All hybrid strains but not the recipient grew indicating that the hybrid strains are *leu* prototroph. (ii) The experiments testing biofilm formation in evolved hybrids have remained inconclusive.

**Fitness measurements.** The selection coefficients reported in Fig. 2a are determined by competition experiments between evolved strains and the ancestral recipient, which are carried out separately in the exponential and stationary growth phase. The selection coefficient between an evolved (hybrid or control) population (*e*) and the ancestor (*a*) in a competition period $t - t_0$ is defined as

$$s_{e,a} = \frac{1}{t-t_0} \log\left(\frac{x_e(t)}{x_a(t)} \Big/ \frac{x_e(t_0)}{x_a(t_0)}\right), \tag{S1}$$

where $x_e(t_0) = 1 - x_a(t_0)$ and $x_e(t) = 1 - x_a(t)$ are the population fractions of the competing lineages at the beginning and at the end of the competition period, respectively. As described in (4), we measure selection coefficients $s_{e,g}$ against a *gfp*-expressing reporter strain and we correct for the fitness effect $s_{g,a}$ of the reporter; i.e., $s_{e,a} = s_{e,g} - s_{g,a}$. Each selection coefficient is obtained as the average over at least 3 independent competition runs. We note that selection coefficients in the exponential phase are essentially differences between cell duplication (birth) rates, whereas stationary-phase selection coefficients are determined by time-dependent birth and death rates. This provides a rationale for the different fitness effects of HGT in the exponential and stationary phase.

## 2. Sequence analysis

**Sequencing pipelines.** DNA-seq reads from each library were properly paired and trimmed in Trimmomatic (version 0.36) (8). We did quality checks of raw and filtered reads using FastQC (v0.11.5)(9).

We used BWA package (v0.7.12-r1039) (10) to align filtered paired reads against the reference genomes (default settings). First, we corrected reference genomes of *B. subtilis* 168 (NCBI, NC_000964.3) and *Bacillus subtilis* subsp. *spizizenii* str. W23 (NCBI, NC_014479.1) for point mutations and short indels present in our ancestral/donor strain with *FastaAlternateReferenceMaker* (GATK, v3.5-0-g36282e4) (11). Adequately, we accommodated all the gene positions in the reference annotation files (gff3, gtf, bed) using custom-made scripts. We mapped filtered paired reads from evolved strains against the adjusted reference genomes using BWA tool (option mem) under two criteria: 1.) loose mapping - a *mismatch penalty* (-B) and *gap open penalty* (-O) of 1 - to determine mutations between the evolved strain and the reference *B. subtilis* 168 genome (NCBI, NC_000964.3), 2.) hard mapping - a *mismatch penalty* (-B) and *gap open penalty* (-O) of 100 - to determine de novo insertions from *Bacillus subtilis* subsp. *spizizenii* str. W23 (NCBI, NC_014479.1) in the evolved species.

We used SAMtools (v1.3.1) (12) to call and pileup SNPs from BWA output files, for each evolutionary run in cycles 9, 15 and 21. We kept only homozygous variants. Based on the quality score distributions for SAM files (Picard v2.6.0) and per base coverage distributions (bedtools, v2.26.0) (13), we filtered out insufficiently covered SNPs and SNPs with poor quality.

**Detection of de novo mutations.** Using the hard mapping pipeline outlined in Sequencing Pipelines, we aligned the reference genomes of the donor and the recipient. We created a master list of positions, and bases, representing point differences between the donor and the recipient genome. For each DNA-seq library in cycle 9, 15 and 21, we called all the mutations that differ in position information, and base composition, compared to the master list.

**Inference of HGT segments.** The ancestral recipient and the donor strain share 3648489 bp of core genome, which is partitioned in 142 segments alternating with segments of the recipient accessory genome (Fig. 1b). The core genome sequence contains 248511 single-nucleotide sites with donor-recipient divergence. At each of these sites, the evolved sequence of a given run has a recipient (R) or donor (D) consensus. We infer transfer segments in the evolved sequence sequentially in the direction of increasing genome coordinate, using the following algorithm (Fig. S1a): (1) The start of each transfer segment is marked by a sequence of two consecutive D alleles (5' marker). The starting coordinate of the segment is then assigned to the midpoint between the 5' marker and its left flanking site, which is either the previous R site or the start of the core genome segment. (2) The end of each segment is marked by a sequence of $k$ consecutive R alleles (3' marker, the optimal value of the parameter $k$ will be inferred below). The end coordinate of the segment is then assigned to the midpoint between the 3' marker and its left flanking site, which is the previous D site.

These rules reflect the fact that the sequence read of an evolved run is determined not only by horizontal transfer, but also by point mutations, mismatch repair, and sequencing errors. These processes introduce local noise to the R/D allele assignment, predominantly singlet allele changes. In previous work (14), transfer segments have been inferred as clusters of D alleles in evolved sequences, whereas our method uses the full local information of R and D alleles.

The results of the transfer inference are robust with respect to variation of the inference algorithm; for example, the distribution of transfer segment lengths and of inter-segment gap lengths depends only weakly on $k$ (Fig. S1b-c; cf. Fig.1d and Fig. S4). This is consistent with most gaps between consecutive transfer segments being much longer than the typical sequence length of 5' and 3' markers.

To further evaluate the performance of the transfer inference algorithm, we record the fraction of R alleles that fall within inferred transfer segments, $x_R = n_R/(n_R + m_R)$, and the fraction of D alleles that fall outside inferred transfer segments, $y_D = m_D/(n_D + m_D)$, where $n_R, n_D$ and $m_R, m_D$ are the allele counts within and outside segments, respectively. In Fig. S1d, the replicate averages of $x_R$ and $y_D$ are plotted for different values of the parameter $k$. Both of these error fractions are small for all $k$, which signals that gene transfer is the dominant process shaping the evolved genomes. Furthermore, the majority (70%) of the remaining R alleles within segments are singlets (i.e., flanked by two D sites), which is consistent with low-rate local point processes independent of gene transfer. We choose $k = 5$ for the analysis of the main text; this parameter value minimizes $x_R + y_D$.

To test the positional accuracy of our inference, we apply the optimized ($k = 5$) algorithm to simulated sequence data representing evolved genomes. These data are generated in two steps: (1) We randomly mark a set $S$ of non-overlapping sequence segments in the recipient core genome, using number and length distribution of the transfer segments recorded in the actual evolved genomes. (2) For all loci with donor-recipient sequence divergence, we randomly assign R/D alleles with probabilities $\tilde{x}_R = n_R/(n_R + n_D)$ and $1 - \tilde{x}_R$ within $S$, and with probabilities $\tilde{y}_R = m_R/(m_R + m_D)$ and $1 - \tilde{y}_R$ outside $S$, respectively. This step preserves the allele frequency distributions recorded in the actual genomes. We then apply the inference algorithm to this artificial genome in order to reconstruct the set $S$. A scatter plot of the inferred length against the input length of $S$ segments shows the high fidelity of the algorithm on individual segments (Fig. S1e). In most cases, an input segment of length $l_{\text{in}}$ is uniquely mapped to an inferred segment of length $l \approx l_{\text{in}}$. In about 0.5% of the cases, two close input segments of length $l_{\text{in},1}$ and $l_{\text{in},2}$ are reported as a paired segment of length $l \approx l_{\text{in},1} + l_{\text{in},2}$; this affects the total coverage only weakly. Defining the inference error $\Delta l = |l - l_{\text{in}}|$ for uniquely mapped segments and $\Delta l = |l - (l_{\text{in},1} + l_{\text{in},2})|$ for paired segments, we find an average value $\Delta l = 18$ bp in the simulated genome data. This uncertainty is of order $d^{-1} = 15$ bp, where $d$ denotes the core genome sequence divergence of the donor and recipient lineages. That is, the positional accuracy of the transfer inference algorithm is set by the density of divergent loci between donor and recipient.

**Inference of de novo insertions.** A separate algorithm was designed to detect the integration of *B. subtilis* W23 specific genes in the evolved replicates. Start and end positions of the accessory *B. subtilis* W23 genes were taken from Table S1 of (15). The genome coverage, per base, was calculated from the stringently mapped reads using bedtools (v2.26.0) (13) and the per base pair depth setting (-d). Segments had to be at least 200 bp long and have an average coverage greater than that of the genome-wide average coverage (ignoring positions with zero mapped reads).

**Inference of deletions.** Using bedtools (version 2.17.0, function coverage) (13), we compute read coverage across the whole genome (window size 100 bp, step by 100 bp) for each DNA-seq library. Genes overlapping with depressed regions (the mean of coverage - 2x standard deviation) are visualized in R packaged TEQC (16) to inspect coverage patterns. Altogether, we find 74 deleted genes. Each of the 74 genes also has zero read counts in a relevant RNA-seq library.

**RNA-seq analysis.** Reads from each library are properly paired and trimmed in Trimmomatic (version 0.38) (8). We map the filtered reads against the reference genome (NCBI, NC_000964.3) adjusted for each evolutionary run (a custom-made script based on the detected and filtered SNPs, avaible at https://github.com/vierocka) using STAR (version 2.5.3a) (17). We call read counts for each known annotated gene using subread packaged (version 1.6.4-Linux-x86_64, tool featureCounts) (18). We normalize the count data using regularized log transformation implemented in R package DESeq2 (19). To evaluate batch effects in sequenced libraries, we use principal component analysis (PCA, R function prcomp). Input for the PCA is a matrix with 30 columns representing sequenced libraries and 4 413 rows representing all detected transcripts. Each cell of the matrix contains a log2-scaled normalized read count. Batch effects can be recognized in the primary libraries as clusters of the PCA components (Fig. S6a).

To remove these effects, we use the sva R package (function combat with parametric adjustments, 2 iterations) (20). The corrected data shows no statistically significant differences between batches (Fig. S6b); this data is reported in Fig. 3 and Table S4.

## 3. Evolutionary analysis of HGT

**Transfer rates of segments.** In the following, we describe our inference procedure for the segment transfer rate, $u(d)$. Here $d$ is the local donor-recipient sequence divergence in a 100 bp window around the recombination start site, which can be at the 5' or 3' terminal end. Because the actual start site for a given transfer segment is unknown, we use a statistical procedure to infer the function $u(d)$. (i) In each of the 595 transfer segments with length > 200 bp found across 7 replicate runs, we record the donor-sequence divergence in 100 bp windows at the 5' and 3' terminal ends. This defines an ordered pair ($d_{\min} = \min(d_{5'}, d_{3'})$, $d_{\max} = \max(d_{5'}, d_{3'})$) for each segment; normalized histograms then yield empirical density distributions $\hat{\rho}_{\min}(d)$ and $\hat{\rho}_{\max}(d)$ for the set of segments (Fig. S3a). We also record a null density $\hat{\rho}_0(d)$ from scrambled 100 bp windows in the core genome. (ii) Guided by previous work(21, 22), we compare this data with a transfer rate model of exponential form,

$$u(d_{rs}|\lambda) = u_0 \exp(-\lambda d_{rs}), \tag{S2}$$

where $d_{rs}$ is the local divergence at the recombination start site, $\lambda$ is the decay exponent, and $u_0$ is a global rate. The parameters $\lambda$ and $u_0$ are to be inferred from the data. In each segment, recombination can start either at the 5' or at the 3' end; this information is a priori unknown. The rate model (S1) defines a divergence probability density

$$\rho_{rs}(d|\lambda) = C \, u(d|\lambda) \, \rho_0(d) \tag{S3}$$

for recombination start sites, where $\rho_0(d)$ is the divergence density in randomly positioned core genome segments and $C$ is a normalization constant. Assuming that the divergence at the recombination end site follows the distribution $\rho_0(d)$, the model predicts density distributions of the minimum and maximum terminal-end distance,

$$\rho_{\min}(d|\lambda) = \rho_{rs}(d|\lambda) \, \Phi_0(d) \tag{S4}$$

$$\rho_{\max}(d|\lambda) = \rho_0(d) \, (1 - \Phi_{rs}(d|\lambda)), \tag{S5}$$

where $\Phi_0(d)$ and $\Phi_{rs}(d|\lambda)$ are the corresponding cumulative distributions. (iii) We calibrate the decay parameter $\lambda$ by a combined least-square fit of the model distributions $\rho_{\min}(d|\lambda)$ and $\rho_{\max}(d|\lambda)$ to the empirical distributions $\hat{\rho}_{\min}(d)$ and $\hat{\rho}_{\max}(d)$ (Fig. S3). The null distribution is estimated from its empirical counterpart, $\rho_0(d) = \hat{\rho}_0(d)$. The good fit of both distributions provides an a-posteriori validation of the exponential rate model given by equation (S2). The divergence-dependent rate $u(d) = u(d|\lambda^*)$ with the optimal decay parameter $\lambda^* = 28$ bp is plotted in Fig. S3b. The global rate $u_0 = 4 \times 10^{-6}/(\text{bp x hr})$ is obtained by calibrating equation (S2) with the total number of observed transfers in the competence periods of the experiment.

**Null model for HGT of genes and operons.** For each gene $g$, we obtain a null probability of transfer, $p_0(g)$, by simulation of a positionally scrambled transfer dynamics with rates $u(d)$ given by equation (S2) with decay parameter $\lambda^* = 28$ bp. First, we build a record of the relative transfer frequencies by simulations of $10^6$ independent single-segment transfers. Each candidate transfer consists of a random starting position in the core genome, a random direction of recombination (i.e., from 5' to 3' end or vice-versa), and a length sampled from the observed length distribution (Fig. 1d). We compute the divergence $d_{\text{rs}}$ in the first 100bp from the recombination start site. We then accept the transfer with probability

$$\pi_0(d_{\text{rs}}) = \tau_0\, u(d_{\text{rs}}), \qquad (S6)$$

where $\tau_0$ is a normalization factor, provided the segment can be contiguously integrated into the recipient core genome (i.e., it does not overlap with an accessory region). For each accepted transfer, we record which genes are affected. This determines probabilities $p_0(g)$ up to the overall normalization $\tau_0$, which is obtained by matching $\sum p_0(g)$ with the average number of genes hit by transfer observed in the data (Fig. 1f). This null model also describes the multiple-hit statistics of transfer in independently evolving replicate populations. Specifically, the distribution of the transfer frequency $\theta$ in $r$ replicate runs is

$$P_0(\theta|p_0) = B(r, \theta r, p_0) \qquad \text{(genes)}, \qquad (S7)$$

where $B(r, k, p_0)$ is the binomial probability for $k$ hits in a sample of size $r$. This distribution has mean $\text{E}(\theta) = p_0$ and variance $\text{Var}(\theta) = r^{-1} p_0 (1 - p_0)$.

The multiple-hit statistics of transfer for operons goes beyond the single-gene model, because it involves spatial correlations between non-overlapping transfer segments in the same replicate run. Here we use a variant of the simulations, where non-overlapping transfer segments are accumulated until the global genome coverage matches the data. For a given operon (23) or PPI community (24), we simulate transfers in the corresponding genome region plus 10000 bp of flanking sequence. We record the mean transfer frequency of the operon, $\bar{\theta} = n_{op}^{-1} \sum_{g \in op} \theta(g)$, where $n_{op}$ denotes the number of genes in the operon and $\theta(g)$ the transfer frequency of each gene. For a given operon ($op$), we obtain the distribution of the mean transfer frequency

$$P_0(\bar{\theta}|op) \qquad \text{(operons)} \qquad (S8)$$

from simulations in batches of $r = 7$ independent replicates. This distribution contains all correlations due to the spatial proximity of genes in the operons. These correlations typically increase the variance of $\bar{\theta}$ compared to the expectation from the single-gene model.

This null model serves for the inference of HGT hot spots and cold spots (Fig. S2 and S4), the GO analysis of HGT (Table S3), and the inference of selection on HGT (Fig. 2c, Fig. 4).

**Inference of hot spots.** We identify HGT hotspots as genome loci containing an operon with significantly enhanced mean transfer frequency, $\bar{\theta} = n_{op}^{-1} \sum_{g \in op} \theta(g)$, compared to the null model described above. We find two significant hot spots: (i) A locus of 7 genes (chromosome positions 2888940 – 2896972) contains the *leu* operon and has $\bar{\theta} = 0.84$. All of these genes are coherently replaced in at least 4 replicates, three of them in all replicates. (ii) A locus of 15 genes (chromosome positions 3514115 – 3529855) contains the two *eps* operons (23) and has $\bar{\theta} = 0.66$. All of these genes are coherently replaced in at least 4 replicates. The statistical significance of these hot spots is evaluated by comparing the observed value of $\bar{\theta}$ with the null distribution (S8) obtained by simulations (Fig. 4e), including a Bonferroni correction for multiple testing in the set of 141 operons with $\geq 5$ genes. We obtain $P < 3 \times 10^{-4}$ for the *leu* locus and $P < 2 \times 10^{-3}$ for the *eps* locus.

**Inference of cold spots.** Cold spots of HGT are genome loci with no observed transfers in any of the replicate runs. We identify statistically significant cold spots from the statistics of gaps between consecutive transfers under the null model of independent transfers. First, in each replicate, the null distribution of gaps is exponential,

$$P_0^{(1)}(\delta) = C\, e^{-\nu_1 \delta} \qquad \text{(gaps)}, \tag{S9}$$

where $C$ is a normalization constant. The empirical gap distribution recorded in the seven replicate runs is consistent with this form with $\nu_1 = 5.22 \times 10^{-5}$/bp for $\delta > 5000$ (Fig. S4a); deviations for short gaps can be attributed to short-range correlations between non-overlapping transfers, constraints imposed by the accessory regions of the genome, and statistical noise in the segment mapping. Next, the distribution of common gaps common in $r$ independent replicates is also exponential,

$$P_0^{(r)}(\delta) = rC\, e^{-r\nu_1 \delta} \qquad \text{(gaps)}. \tag{S10}$$

The observed distribution of common gaps in 7 replicate runs is asymptotically exponential, $\hat{Q}^{(7)}(\delta) = C_7\, e^{-\nu_7 \delta}$ with an exponent $\nu_7 = 1.88 \times 10^{-4} < 7\nu_1$ for $\delta > 5000$ bp (Fig. S4b). This inequality reveals weak correlations between replicates, which are in tune with the genome-wide selection model discussed below. We find two significant cold spots as outliers with respect to $\hat{Q}^{(7)}(\delta)$. The first cold spot has a length of 48815bp (chromosome positions 332356 – 381171) and contains 43 genes; the second has a length of 55895bp (chromosome positions 1823905 – 1879800) and contains 28 genes, including the ultra-long genes *pksN* (16467bp) and *pksR* (7632bp). We evaluate the significance of the cold spots from the Gumbel distribution (parameters 33839, 5409) obtained from $\hat{Q}^{(7)}(\delta)$ for 477 draws, corresponding to the number of gaps common to all replicates. We obtain $P < 0.041$ for the first and $P < 0.01$ for the second cold spot.

**GO analysis.** We study the association of HGT with cellular functions, using GO categories obtained from (25). For each category containing at least 10 genes, we evaluate the mean transfer frequency $\bar{\theta} = n_{\text{cat}}^{-1} \sum_{g \in \text{cat}} \theta(g)$. We compare this quantity with its expectation value under the null model, $E(\bar{\theta}) = \bar{p}_0 = n_{\text{cat}}^{-1} \sum_{g \in \text{cat}} p_0(g)$, by evaluating the ratio $\varphi = \bar{\theta}/\bar{p}_0$ (Table S3). Because our null model incorporates variations in local sequence similarity, we discount correlations between GO categories and HGT generated by their mutual dependence on sequence similarity. To evaluate the statistical significance of over- or underrepresentation of HGT in a given category, we compare the observed number of transfer hits in 7 replicates, $t_{\text{cat}} = 7\, n_{\text{cat}}\, \bar{\theta}$, with the expected number, $E(t_{\text{cat}}) = 7\, n_{\text{cat}}\, \bar{p}_0$, using a standard Binomial test with a Bonferroni multiple-testing correction in the set of GO categories (Table S3).

**Selection model for HGT of genes.** Selection affects the probability to observe the transfer of an individual gene $(g)$,

$$p(g) = p_0(g)\, \varphi\big(2N_e s(g)\big). \tag{S11}$$

Here $p_0(g)$ is the transfer probability under the null model; $\varphi$ is the ratio of fixation probabilities under the full model and under the null model for genetic variants with orthologous replacement of the gene. The function $\varphi(2N_e s(g))$ depends on a gene-specific selection coefficient of transfer, $s(g)$, and the effective population size $N_e$ underlying the population dynamics in the evolution experiments (26). For the inference of genome-wide selection acting on HGT, we determine $p_0(g)$ by simulations, as described above. Next, we partition the genome into two classes with uniform selection in each class, i.e., $\varphi(g) = \varphi_+$ in the (+) class and $\varphi(g) = \varphi_-$ in the (-) class. This assumption is expected to dilute the gene-specific signal of selection, but is necessary to avoid overfitting of the selection model parameters from the data. The two-class model has a transfer probability

$$p(c, \varphi_+, \varphi_-, p_0) = c\, \varphi_+\, p_0 + (1-c)\, \varphi_-\, p_0, \tag{S12}$$

where the mixture parameter $c$ measures the fraction of core genes in the (+) class. The resulting distribution of the transfer frequency $\theta$ in $r$ replicate runs,

$$Q(\theta|c, \varphi_+, \varphi_-, p_0) = c\, B(r, \theta r, \varphi_+ p_0) + (1-c)\, B(r, \theta r, \varphi_- p_0), \tag{S13}$$

is to be compared with the corresponding probability under the null model, which is given by equation (S7).

**Bayesian inference of selection on HGT.** We infer the mixture model parameters $c, \varphi_+, \varphi_-$ using the log likelihood score

$$S(c, \varphi_+, \varphi_-) = \sum_g \log \frac{Q(\theta(g)|c, \varphi_+, \varphi_-, p_0(g))}{P_0(\theta(g)|p_0(g))} \tag{S14}$$

with probability distributions $Q$ and $P_0$ given by equations (S7) and (S13). Evaluating this score function in our data set with flat prior probabilities of $(c, \varphi_+, \varphi_-)$ leads to the posterior score distribution shown in Fig. S6a. Varying the parameter $c$, we evaluate the conditional score maximum $S^*(c)$ (Fig. S5b) and the corresponding parameters $\varphi_+^*(c), \varphi_-^*(c)$ (Fig. S5c). This determines the absolute maximum-likelihood score $S^* = \max_q S^*(c) = 47$ and the maximum-likelihood parameters $(c^*, \varphi_+^*, \varphi_-^*) = \arg\max_{c, \varphi_+, \varphi_-} S(c, \varphi_+, \varphi_-) = (0.2, 1.9, 0.75)$. We estimate the parameter regime of high posterior probability by integration of the function $S^*(c)$. Using conservative confidence thresholds (90% for $c < c^*$, 50% for $c > c^*$), we obtain

$$c \in [0.1, 0.4], \; \varphi_+ \in [1.6, 2.4], \; \varphi_- \in [0.6, 0.84]. \tag{S15}$$

The expected fraction of (+) genes can also be evaluated for each transfer class. Specifically, in the high-transfer (low-transfer) class defined in the main text, 88% (20%) of the genes are inferred to be under positive selection. This calculus also yields an estimate of the fraction of observed transfers that are under positive selection,

$$\eta = \frac{\varphi_+ c}{\varphi_+ c + \varphi_-(1-c)}. \tag{S16}$$

We obtain $\eta = 0.40\ [0.20, 0.60]$; the $c$-dependent conditional fraction $\eta^*(c)$ is shown in Fig. S5d. To estimate the statistical significance of this inference, we rewrite the score (S14) in the form

$$S = n\, H(Q_2 \mid P_{0,2}), \tag{S17}$$

where $n$ is the number of core genes and $H(Q_2 \mid P_{0,2})$ is the relative entropy (Kullback-Leibler distance) between the joint distributions $Q_2(\theta, p_0) = Q(\theta|p_0) P(p_0)$ and $P_{0,2}(\theta, p_0) = P_0(\theta|p_0) P(p_0)$ (here we omit the parameters $c, \varphi_+, \varphi_-$ from the notation). At the maximum-likelihood point, Sanov's theorem then gives the probability of the observed data under the null model, $P < \exp(-S^*) \sim 10^{-20}$.

To display the deviations of the transfer statistics from the neutral null model, Fig. 3b shows the empirical likelihood ratio $\hat{Q}(\theta|p_0)/P_0(\theta|p_0)$ in four bins of $p_0$. In Fig. 2c, we plot the likelihood ratio $Q^*(\theta)/P_0(\theta)$, where $Q^*(\theta) = \int Q_2^*(\theta, p_0)\, dp_0$ is the marginal distribution of the maximum-likelihood selection model and $P_0(\theta) = \int P_{0,2}(\theta, p_0)\, dp_0$ is the marginal distribution of the null model. Comparison with the corresponding empirical ratio $\hat{Q}(\theta)/P_0(\theta)$ shows that the maximum-likelihood selection model provides a good fit to the data, whereas the null model does not. We emphasize that our inference of selection and significance estimate are based on the full joint data $(\theta, p_0)$, in order to properly discount local variations of $p_0$.

Importantly, the null model given by equation (S6) takes into account local variations of the sequence divergence between donor and recipient, avoiding a spurious signal of selection from this variation. Our inference of positive and negative selection is based on the assumption that this null model is a faithful representation of the neutral transfer dynamics. However, even if the null model contains some uniform selection on HGT, our inference still provides evidence of two gene classes with selective differences ($\varphi_+ > \varphi_-$).

**HGT in operons.** Our analysis of HGT in gene networks uses a list of operons in *B. subtilis* retrieved from (23). To infer selection on HGT for a given operon (*op*), we compare the observed mean transfer frequency of the network genes, $\bar{\theta} = n_{op}^{-1} \sum_{g \in op} \theta(g)$, with the neutral null distribution $P_0(\bar{\theta}|op)$ given by equation (S8).

For the aggregate inference of negative selection on operons (Fig. 4a), we use the set of all *B. subtilis* operons in the core genome region that contain at least 7 genes; the threshold is chosen to ensure any signal of selection can be considered a network property (i.e., is not dominated by single genes). This set contains 56 operons with an average of 10 genes per operon. Compared to the neutral null model, the data shows overrepresentation of transfers at low frequency, 46% of counts of $\bar{\theta} < 0.1$ in the data vs. 31% in the null model. These deviations are statistically significant, signaling selection against HGT at the network level ($< 10^{-3}$, computed again by Sanov's theorem from the histogram of data and the null distribution). We have excluded the hotspots for recombination (*leu* and *eps* operons), which are analyzed separately for positive selection.

The individual operons *leu* and *eps* differ significantly from the null model ($P < 3 \times 10^{-4}$ for *leu*, $P < 2 \times 10^{-3}$ for *eps*; see above). The additional examples of high- and low-transfer operons (Fig. 4d, Fig. S7) are marginally significant at the individual level ($P < 0.05$ without stringent Bonferroni correction). The list of genes in these operons is given in Table S5.

**HGT dynamics in a cross-lineage fitness landscape.** Intra-network selection on HGT can be described by a minimal cross-lineage fitness landscape $F(q|a, b)$ that depends on the fraction of transferred network genes, $q$, as given by equation [2]. Simulations of the HGT dynamics in the landscape $F(q|a,b)$ use a transfer probability for segments,

$$\pi(d_{rs}, q, q'|a,b) = \pi_0(d_{rs}) \, \varphi\big(2N_e \, s(q, q'|a,b)\big), \tag{S18}$$

where $q, q'$ are the network transfer fractions before and after the event in question, and $\pi_0(d_{rs})$ is the neutral transfer probability given by equation (S6). The fixation probability $\varphi$ depends on the effective population size $N_e$ and on the selection coefficient $s(q, q'|a,b) = F(q'|a,b) - F(q|a,b)$; cf. equation (S11). These dynamics determine distributions of the mean transfer frequency $Q(\bar{\theta}|a,b)$ in the fitness landscape $F(q|a,b)$, which is to be compared with the corresponding null distribution $P_0(\bar{\theta})$ given by equation (S8). The model distributions $Q(\bar{\theta}|a,b)$ reported in Fig. 4a are obtained from the fitness model of equation [2] with scaled parameters $2N_e a = -0.4$, $2N_e b = 0$ (negative directional selection, orange) and $2N_e a = 0$, $2N_e b = -2.2$ (hybrid incompatibilities, blue).

The distributions of Fig. 4d (*yes* and *iol*) have parameters $2N_e b = 0$ and $2N_e a = -10$ (negative directional selection). The distributions of Fig. 4e and Fig. S7 have parameters $2N_e a = 5.4$ (*leu*), $2N_e a = 3.5$ (*eps*), $2N_e a = 1.5$ (*bkd*), $2N_e a = 1.4$ (*xse*), and $2N_e b = 0$ (positive directional selection, magenta). These selection coefficients are calibrated such that the model distributions $Q(\bar{\theta}|a,b)$ have the same average of $\bar{\theta}$ as the histogram of operon data. The analysis shows that the observed HGT in operons is broadly consistent with selection given by a fitness landscape of the form of equation [2]. However, more data is needed to calibrate mixture models with maximum-likelihood parameters delineating directional and epistatic selection.

**Data availability.** Data is available in Table S1 (de novo mutations), Table S2 (genomic positions of orthologous recombinations, insertions from accessory genome of donor, deletions, duplications), Table S3 (GO categories), Table S4 (log$_2$ fold changes $\Delta R$ of mRNA levels), Table S5 (operons with constituent genes). All custom-made scripts will be available at https://github.com/vierocka.

**Supporting Figures and Tables**

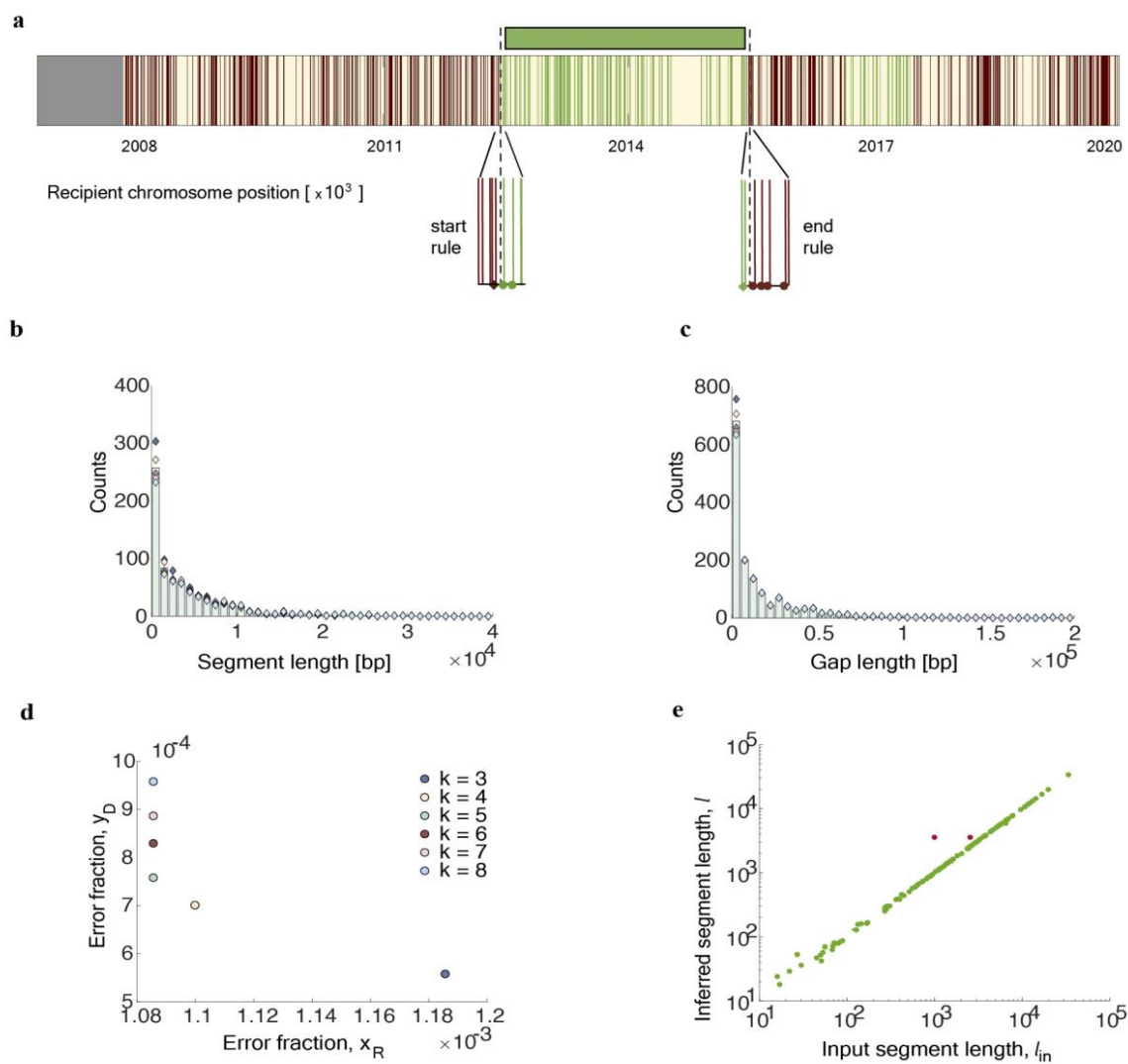

**Fig. S1. Inference of orthologous recombination. (a)** Algorithm. The recipient genome is partitioned into core (yellow) and accessory (grey) segments. Core sequence sites with donor-recipient divergence are marked by vertical lines. The evolved sequence has recipient (R, red) or donor (D, green) alleles at each of these sites (data shown from run R1, genomic coordinates 2006770 – 2020000). The start of each segment is determined by a 5' marker (two consecutive D alleles, green dots); the end is determined by a 3' marker ($k = 5$ consecutive R alleles, red dots). These markers and their left flanking sites (diamonds) set the start/end coordinates of inferred transfer segments (green bars) by a midpoint rule. See SI Appendix for details. **(b,c)** Length distribution of inferred transfer segments and of inter-segment gaps for different values of the parameter k. **(d)** Fraction of R alleles that fall within transfer segments, $x_R$, and fraction of D alleles that fall outside inferred transfer segments, $y_D$, for different values of the parameter k. **(e)** Test of the algorithm on simulated data. Scatter plot of the inferred length vs. the input length for a set of 117 scrambled transfer segments. This set produces 115 inferred segments uniquely mapped to an input segment (green dots) and one inferred segment containing two close input segments (red dots). The average inference error is $\Delta l = 20$ bp. See section 2 of SI Appendix for details.

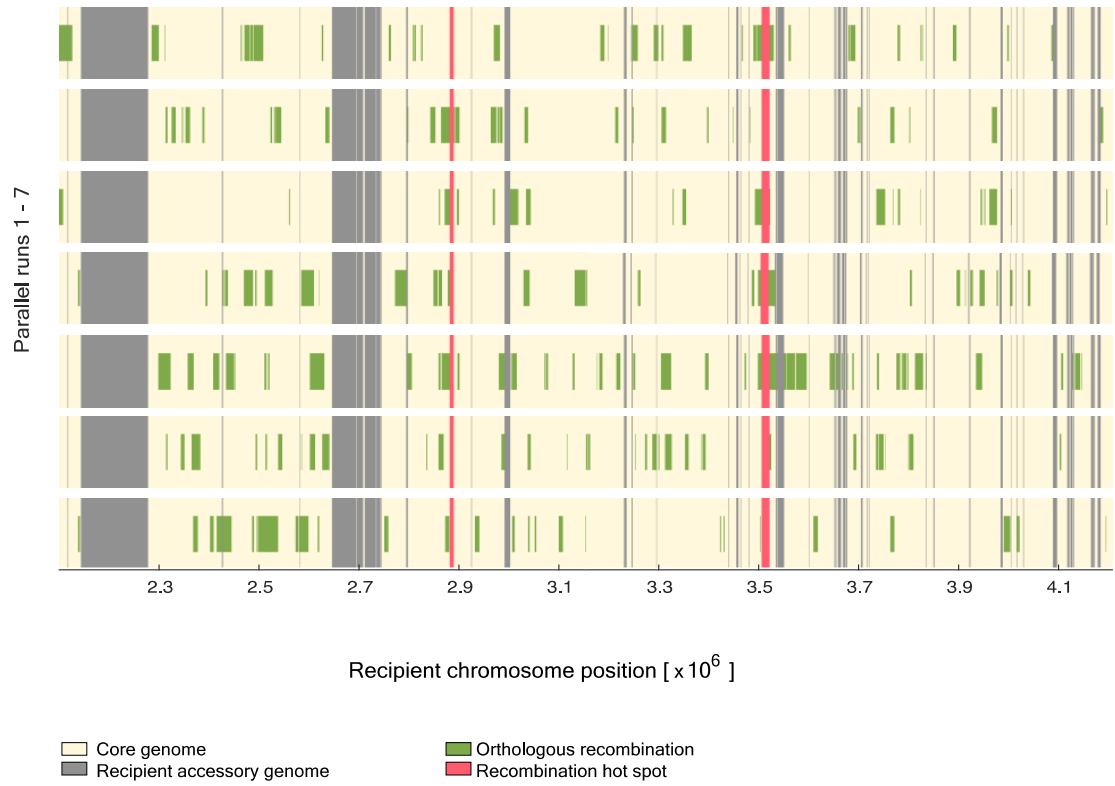

**Fig. S2. Genomics of HGT across replicate runs.** The aligned HGT pattern in all 7 parallel runs is shown for a part of the genome (annotation as in Fig. 1b).

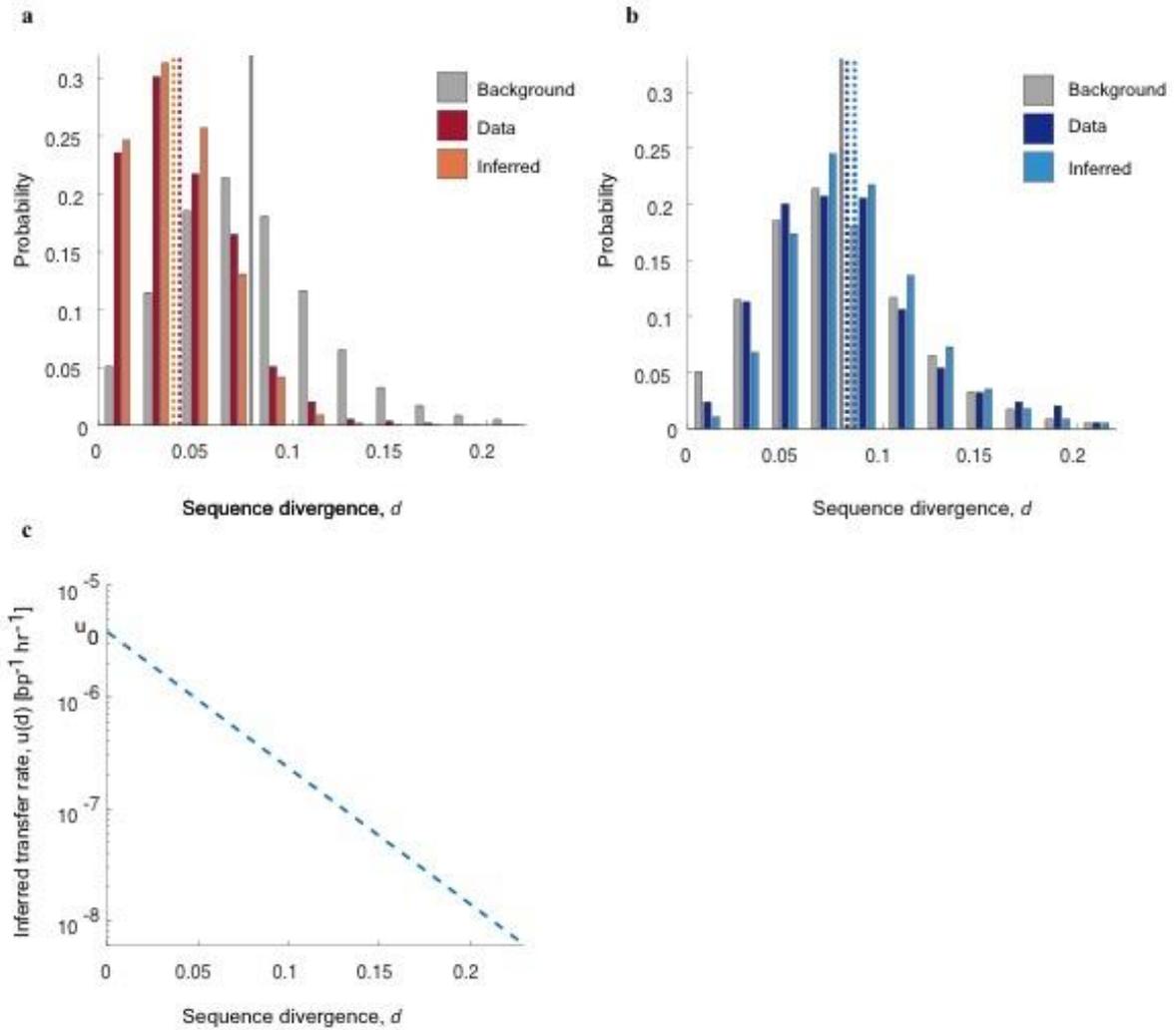

**Fig. S3. Transfer rate depends on local sequence divergence.** (**a**) Observed sequence divergence $d$ in 100bp terminal regions of transferred genome segments: distribution $\hat{\rho}_{\min}(d)$ recorded from the low-divergence end within each segment (left panel, red) and (**b**) distribution $\hat{\rho}_{\max}(d)$ recorded from high-divergence ends (right panel, blue). These data are shown together with the background distribution $\hat{\rho}_0(d)$ obtained from scrambled 100 bp windows (gray) and with the distributions $\rho_{\min}(d)$ and $\rho_{\max}(d)$ computed from the optimal local-divergence model. Dashed lines mark the mean values of the corresponding distributions. (**c**) Inferred transfer rate depending on the local sequence divergence at the recombination start site, $u(d) = u_0 \exp(-\lambda d)$ with $\lambda = 28$ bp and $u_0 = 4 \times 10^{-6}/(\text{bp} \times \text{hr})$. See section 3 of SI Appendix for definitions of these distributions and for details of the inference procedure.

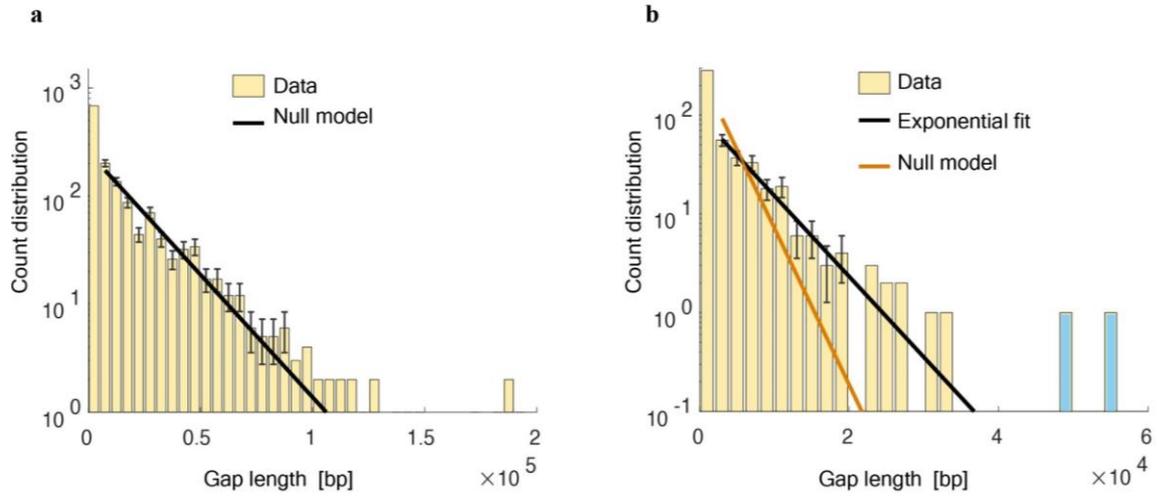

**Fig. S4. Inference of HGT cold spots.** (**a**) Distribution of the gap length δ between subsequent transfers in single runs with exponential fit, $P_0^{(1)}(\delta) \sim \exp(-\nu_1 \delta)$ for $\delta > 3000$ bp (black line). (**b**) Length distribution of common gaps in all 7 runs together with exponential fit, $Q^{(7)}(\delta) \sim \exp(-\nu_7 \delta)$ (black line), shown together with null expectation for independent replica, $P_0^{(7)}(\delta) \sim \exp(-7\nu_1 \delta)$ (orange line). Two cold spots of length $l = 48815$ bp and $l = 55895$ bp (blue) are identified as statistically significant outliers to the observed exponential pattern ($P < 0.04$). See section 3 of SI Appendix for the statistical analysis.

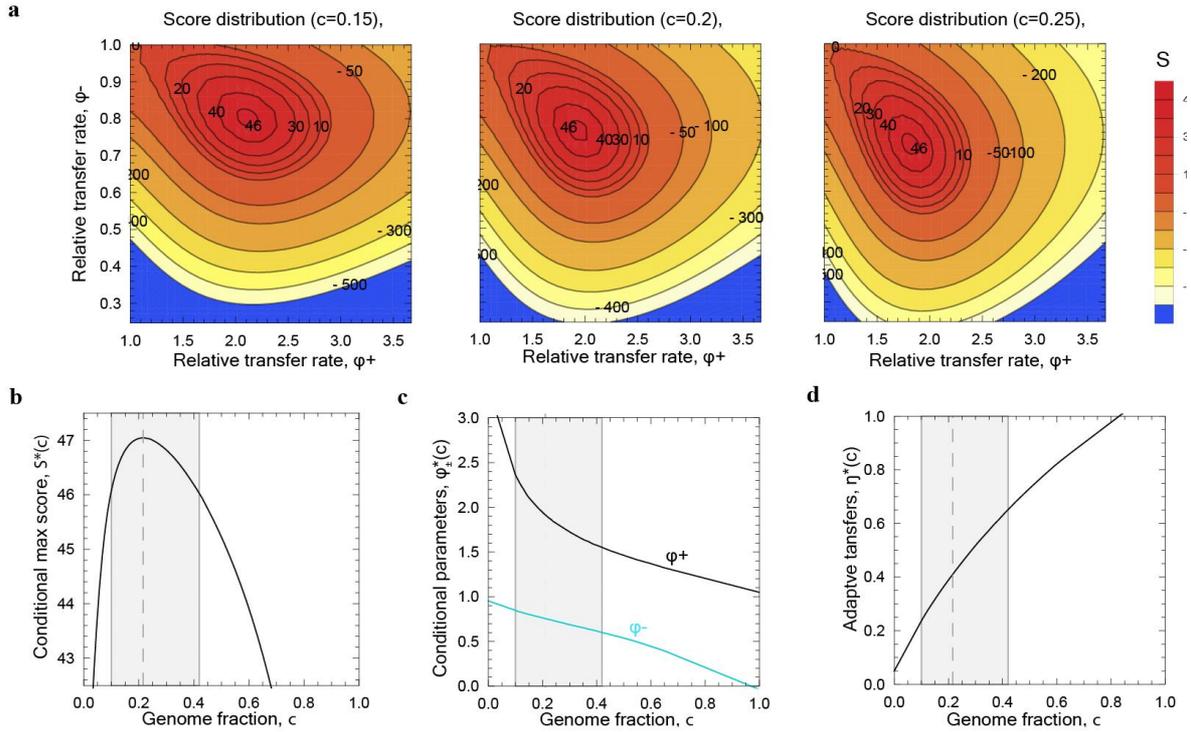

**Fig S5. Inference of the selection model for transfer.** (**a**) Log likelihood score of the selection model compared to the neutral null model as a function of the relative transfer rates $\varphi_+ = p_+/p_0$ and $\varphi_- = p_-/p_0$ for given genome fractions $c = 0.15, 0.25, 0.35$. (**b**) Conditional maximum log likelihood score $S^*(c)$. The $c$ interval of high posterior probability is indicated by shading. (**c**) Conditional maximum-likelihood parameters $\varphi_+^*(c)$ and $\varphi_-^*(c)$. (**d**) Partitioning of HGT by selection class. Conditional maximum likelihood function $\eta^*(c)$, where $\eta = \varphi_+ c/(\varphi_+ c + \varphi_-(1-c))$ is the fraction of the observed transfers predicted to be under positive selection. See section 3 of SI Appendix for details of the inference procedure.

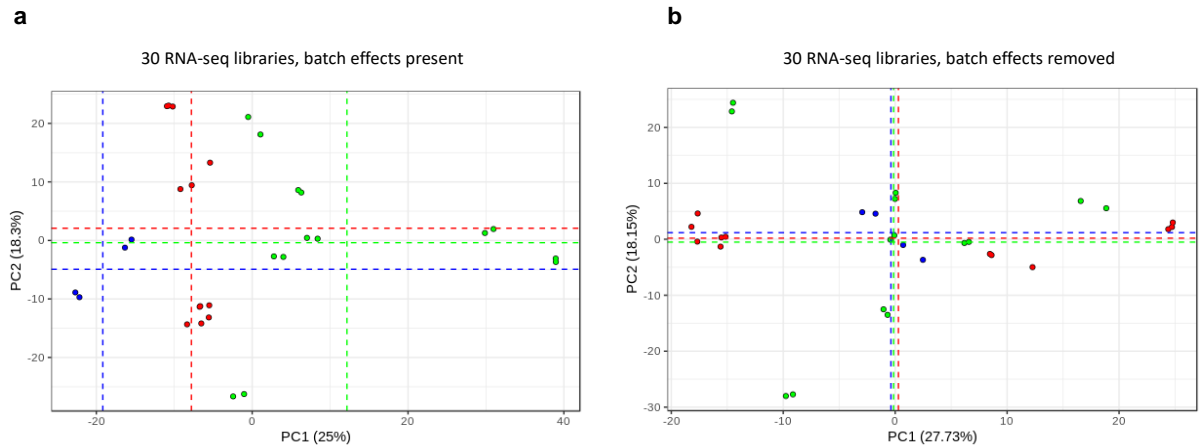

**Fig. S6. Batch correction of transcriptomics data. (a)** 30 RNAseq libraries from 7 evolved populations, 3 control runs, and the ancestral recipient were obtained in three batches (red, blue, green). We perform a principal component analysis (PCA) of the $\log_2$ -scaled normalized read counts of these libraries; batch averages of the first two PC components are shown as dashed lines. Batch effects can be inferred primarily in the component PC1. **(b)** After batch correction (SI Appendix), the data show no longer statistically significant differences between batches.

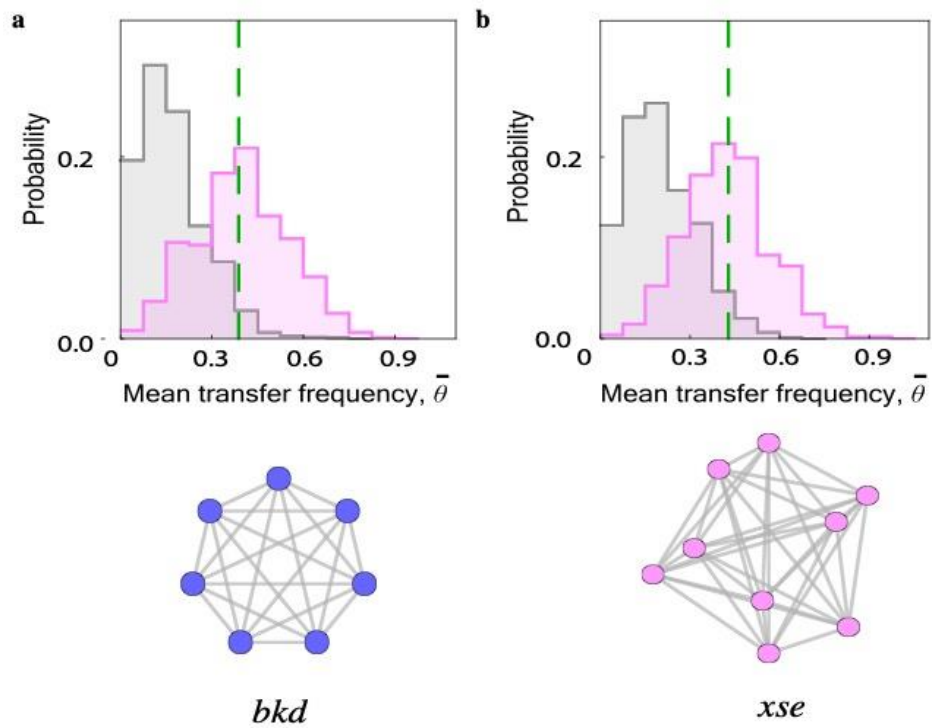

**Fig S7. Additional operons with enhanced HGT**. (**a**) *bkd* operon, with 7 genes and $\bar{\theta} = 0.39$. (**b**) *xse* operon with 9 genes and $\bar{\theta} = 0.43$. The observed transfer frequency $\bar{\theta}$ (green, dashed line) is substantially enhanced in comparison to the neutral null model (gray) and consistent with positive directional selection ($a > 0$, magenta). Both operons are associated to metabolism of amino acids and small compounds (24). The corresponding PPI networks are shown below.